\documentclass[prc,twocolumn,showpacs,preprintnumbers,amsmath,amssymb,superscriptaddress]{revtex4}

\usepackage{graphicx} % Include figure files
\usepackage{dcolumn}  % Align table columns on decimal point
\usepackage{bm}       % bold math

\newcommand{\be}{\begin{equation}}
\newcommand{\ee}{\end{equation}}

\newcommand{\hfbax}{\sc hfb-ax}
\newcommand{\hfbrad}{\sc hfbrad}
\newcommand{\hfbtho}{\sc hfbtho}
\newcommand{\hfbvan}{\sc hfb-2d-lattice}

\newcommand{\Vvec}{\boldsymbol V}
\newcommand{\Fvec}{\boldsymbol F}
\newcommand{\Ivec}{\boldsymbol I}

\begin{document}

\title{Deformed Coordinate-Space  Hartree-Fock-Bogoliubov Approach to
 Weakly Bound Nuclei and Large Deformations}

\author{J.C. Pei}
%\email{peij@ornl.gov}
\affiliation{Joint Institute for Heavy Ion Research, Oak Ridge, TN
37831, USA}
\affiliation{Department of Physics and
Astronomy, University of Tennessee, Knoxville, TN 37996, USA}
\affiliation{Oak Ridge National Laboratory, P.O. Box 2008, Oak
Ridge, TN 37831, USA}
\author{M.V. Stoitsov}
\affiliation{Department of Physics and Astronomy, University of
Tennessee, Knoxville, TN 37996, USA}
\affiliation{Oak Ridge National
Laboratory, P.O. Box 2008, Oak Ridge, TN 37831, USA}
\affiliation{Institute of Nuclear Research and Nuclear Energy,
Bulgarian Academy of Sciences, Sofia, Bulgaria}
\author{G.I. Fann}
\affiliation{Oak Ridge National Laboratory, P.O. Box 2008, Oak
Ridge, TN 37831, USA}
\author{W. Nazarewicz}
\affiliation{Department of Physics and Astronomy, University of
Tennessee, Knoxville, TN 37996, USA}
\affiliation{Oak Ridge National
Laboratory, P.O. Box 2008, Oak Ridge, TN 37831, USA}
\affiliation{Institute of Theoretical Physics, Warsaw University,
ul.Ho\.{z}a 69, PL-00681 Warsaw, Poland}
\author{N. Schunck}
\affiliation{Department of Physics and Astronomy, University of
Tennessee, Knoxville, TN 37996, USA}
\affiliation{Oak Ridge National
Laboratory, P.O. Box 2008, Oak Ridge, TN 37831, USA}
\author{F.R. Xu}
\affiliation{State Key Laboratory of Nuclear Physics and Technology,
School of Physics, Peking University, Beijing 100871, China}

\date{\today}

\begin{abstract}

The coordinate space formulation of the Hartree-Fock-Bogoliubov
(HFB) method enables self-consistent treatment of mean-field and
pairing in  weakly bound systems whose properties are affected by
the particle continuum space. Of particular interest are
neutron-rich, deformed drip-line nuclei which can exhibit novel
properties associated with neutron skin. To describe such systems
theoretically, we developed an accurate 2D lattice  Skyrme-HFB
solver {\hfbax} based on B-splines. Compared to previous
implementations, we made a number of improvements aimed at boosting
the solver's performance. These include: explicit imposition of
axiality and space inversion, use of the modified Broyden's method
to solve self-consistent equations, and a partial parallelization of
the code. {\hfbax} has been benchmarked against other HFB solvers,
both spherical and deformed, and the accuracy of the B-spline
expansion was tested by employing the multiresolution wavelet
method. Illustrative calculations are carried out  for stable and
weakly bound nuclei at spherical and very deformed shapes, including
constrained  fission pathways. In addition to providing new physics
insights, {\hfbax} can serve as a useful tool to assess the
reliability and applicability  of coordinate-space and
configuration-space HFB solvers, both existing and in development.

\end{abstract}

\pacs{21.60.Jz, 21.10.-k, 02.70.-c, 31.15.ej}

\maketitle

\section{Introduction}

The important new focus in theoretical nuclear structure research is
to develop a coherent  theoretical framework aiming at the
microscopic description of nuclear many-body  systems and capable of
extrapolating into unknown  regions. An important component in the
theoretical landscape, and a crucial part of the theory roadmap
\cite{RIAtheory,unedf}, is the nuclear Density Functional Theory
(DFT) in the formulation of Kohn and Sham \cite{KS}. Since the
majority of  nuclei in their ground states are superconductors,
pairing correlations have to be taken into account. The resulting
HFB or Bogoliubov de-Gennes equations can be viewed as the
generalized Kohn-Sham equations of the standard DFT. The main
ingredient of the nuclear DFT \cite{DFTrev,DFTrev1} is the energy
density functional that depends on proton and neutron densities and
currents, as well as pairing densities representing correlated
nucleonic pairs \cite{LDA}.

The unique  structural factor that determines many properties of weakly
bound nuclei is the closeness of
 the particle continuum. While the
nuclear densities of bound nuclei eventually vanish at large
distances, the  wave functions  of positive-energy states do not
decay outside the nuclear volume, and this can be a source of
significant theoretical difficulties. This problem is naturally
overcome in the HFB  method with a realistic pairing interaction in
which the coupling of bound states to the particle continuum is
correctly taken into account. In this context, particularly useful
is the coordinate-space formulation of HFB
\cite{dobaczewski84,dobaczewski96}. Another advantage of this method
is its ability to treat arbitrarily complex intrinsic shapes,
including those seen in fission or fusion.

The HFB equations of the nuclear DFT represent the self-consistent
iterative convergence schemes. Their  computational cost  can become
very expensive, especially when the size of the model space --
largely determined by the self-consistent symmetries imposed --  or
the number of nuclear configurations processed simultaneously is
huge. The advent of teraflop computers  makes such large-scale
calculations feasible, but in order to better optimize unique
resources,  new-generation tools are called for.

A number of coordinate-space approaches to the nuclear DFT have been
developed over the years, and their performance strongly depends on
the size and symmetries of the spatial mesh employed
\cite{blum,Ter96,heenen1}. The {\sc ev8} code  solves the
Hartree-Fock plus BCS equations for Skyrme-type functionals via a
discretization of the individual wave-functions on a 3D Cartesian
mesh \cite{bonche}, assuming three symmetry planes. The  1D
{\hfbrad} finite-difference code
 has been developed as a standard tool for HFB
calculations of spherical nuclei~\cite{bennaceur}. While limited to
the radial coordinate only, {\hfbrad} allows very precise
calculations, as the mesh step can be taken very low. The recently
developed parallel 2D {\hfbvan} code
~\cite{teran,oberacker,blazkiewicz} is based on B-splines; it can
treat axially deformed nuclei, {\it including} those with
reflection-asymmetric shapes. No coordinate-space 3D HFB nuclear
framework exists at present; however, a number of developments are
under way, including a general-purpose HFB solver based on wavelet
technology \cite{madness,fann}.

One traditional way of solving the HFB problem is to use the configuration-space
technique, in which  HFB eigenstates are expanded
in a discrete basis, such as the
harmonic oscillator (HO) basis \cite{ring80}.
 This method is very efficient and
has been applied successfully in the large-scale calculations of
nuclear properties~\cite{dobaczewski04}. However, the use of the HO
basis is questionable in the limit of weak binding (because of
incorrect asymptotic behavior of HO states) and at very large
deformations, both requiring the use of unrealistically large
configuration spaces. In both situations, a coordinate-state
description is superior.

The main objective  of this paper is to develop a reliable and
accurate HFB axial solver, based on  B-splines, to study weakly
bound nuclei  and/or constrained energy surfaces involving very
large deformations. An HFB scheme based on such a concept  has been
implemented in {\hfbvan} by the Vanderbilt group
~\cite{teran,oberacker,blazkiewicz}.  An attractive feature  of
{\hfbvan}  is that by taking high-order B-splines one guarantees the
correct representation of derivative operators on the spatial
lattice \cite{blum}. Unfortunately, the performance of {\hfbvan} is
strongly CPU-limited. To speed up the  calculations, and at the same
time to improve the accuracy, we wrote a new 2D HFB B-spline code,
called {\hfbax}, in which a number of new features were introduced.
Firstly, we incorporated space inversion as a self-consistent
symmetry. Since reflection-asymmetric ground-state deformations are
present only in a handful of nuclei, this is not a serious
limitation. Furthermore, we improved  the iterative algorithm by
means of the modified Broyden's method. This resulted in significant
convergence acceleration. We also made a number of smaller
optimizations to the HFB solver which are described in the text. The
performance and accuracy of  {\hfbax} has been very carefully tested
against other codes. In short, we developed  a fast, deformed
coordinate-space HFB solver that can be used to carry out
large-scale calculations on leadership-class computers, and it can
also be invaluable when benchmarking the next-generation nuclear DFT
tools.

This paper is organized as follows. Section~\ref{theoret} shortly
summarizes the coordinate-space HFB approach in the cylindrical
coordinate system and  describes the numerical method used.  In Sec.~\ref{benchmarking},
numerical tests are presented, with an emphasis on benchmarking
against other existing HFB solvers. The examples include: (i) a
two-center potential problem and a  comparison with the multiresolution
wavelet method and the HO expansion technique; (ii) comparison with the spherical
finite-difference solver {\hfbrad} for stable $^{120}$Sn and drip-line
Ni isotopes; (iii) study of neutron-rich deformed $^{102,110}$Zr and
comparison with the axial solvers {\hfbtho}
and {\hfbvan}; and (iv) fission path
calculations for  $^{240}$Pu. Finally, Sec.~\ref{conclusions} contains
the main conclusions of this paper.

\section{Theoretical framework and numerical method}\label{theoret}

\subsection{HFB equation in cylindrical coordinate space}

The HFB equations in the coordinate-space representation
can be written as~\cite{dobaczewski84,dobaczewski96,stoitsov05}:
\begin{equation}\label{HFBm}
\begin{array}{c}
\displaystyle \int d\mathbf{r'} \displaystyle \sum_{\sigma'} \left(
\begin{array}{cc}
h(\mathbf{r}\sigma,\mathbf{r'}\sigma')-\lambda&\tilde{h}(\mathbf{r}\sigma,\mathbf{r'}\sigma') \vspace{3pt} \\
\tilde{h}(\mathbf{r}\sigma,\mathbf{r'}\sigma') &-h(\mathbf{r}\sigma,\mathbf{r'}\sigma')+\lambda\\
\end{array}
\right) \vspace{5pt} \\   ~~~~~~~~~~ \times\left(
\begin{array}{c}
\psi^{(1)}(\mathbf{r'}\sigma') \vspace{2pt}\\
\psi^{(2)}(\mathbf{r'}\sigma') \\
\end{array}
\right)=E\left(
\begin{array}{c}
\psi^{(1)}(\mathbf{r}\sigma) \vspace{2pt}\\
\psi^{(2)}(\mathbf{r}\sigma) \\
\end{array}
\right),
\end{array}
\end{equation}
where  ($\mathbf{r}, \sigma$) are the particle spatial and spin
coordinates, $h(\mathbf{r}\sigma,\mathbf{r'}\sigma')$ and
$\tilde{h}(\mathbf{r}\sigma,\mathbf{r'}\sigma')$ are the
particle-hole (p-h) and particle-particle (p-p)
  components of the
single-quasiparticle Hamiltonian, respectively,
$\psi_n^{(1)}(\mathbf{r}\sigma)$ and
$\psi_n^{(2)}(\mathbf{r}\sigma)$ are the upper and lower components
of the single-quasiparticle HFB wave function, and $\lambda$ is the
chemical potential. The spectrum of quasiparticle energies $E$ is
discrete for $|E|<-\lambda$ and continuous for $|E|>-\lambda$. With
the box boundary conditions, the continuum is discretized. In
practical calculations, the p-h channel is often modeled with the
Skyrme energy density functional, while  a zero-range $\delta$
pairing interaction is used in the p-p channel. This choice is
motivated by the fact that zero-range interactions yield the local
HFB equations in coordinate space which are easy to solve.

In the axially symmetric geometry, the third component of the
single-particle  angular
momentum, $\Omega$,  is a good quantum number.
The HFB wave function can thus be written as $\Psi_n(\mathbf{r},\Omega,
q)$ where $\mathbf{r}$=($\phi$, $\rho$, $z$), $q$=$\pm\frac{1}{2}$
denotes the cylindrical
isospin coordinates, and $\Omega$=$\pm\frac{1}{2}$,
$\pm\frac{3}{2}$, $\pm\frac{5}{2}$, $\ldots$.
The corresponding HFB wave function can be written as~\cite{teran},
\begin{equation}\label{wf4}
\begin{array}{ccl}
\Psi_{n}(\mathbf{r},\Omega,q)&=&\left(
\begin{array}{c}
\psi_{n,\Omega,q}^{(1)}(\phi, \rho, z) \vspace{2pt} \\
\psi_{n,\Omega,q}^{(2)}(\phi, \rho, z)\\
\end{array}
\right) \vspace{5pt} \\
 &=&\displaystyle \frac{1}{\sqrt{2\pi}}\left(
\begin{array}{c}
e^{i(\Omega-\frac{1}{2})\phi}\psi_{n,\Omega,q}^{(1)}(\rho,z, \uparrow)\vspace{1pt} \\
e^{i(\Omega+\frac{1}{2})\phi}\psi_{n,\Omega,q}^{(1)}(\rho,z, \downarrow)\vspace{1pt} \\
e^{i(\Omega-\frac{1}{2})\phi}\psi_{n,\Omega,q}^{(2)}(\rho,z, \uparrow)\vspace{1pt} \\
e^{i(\Omega+\frac{1}{2})\phi}\psi_{n,\Omega,q}^{(2)}(\rho,z, \downarrow)\\
\end{array}
\right). \\
\end{array}
\end{equation}

Following the notation of
Ref.~\cite{teran}, we introduce
\begin{equation}
\begin{array}{c}
 \displaystyle U_{n\Omega q}^{(1,2)}(\rho, z)=\psi_{n\Omega q}^{(1,2)}
 (\rho, z, \uparrow), \vspace{5pt}\\
\displaystyle D_{n\Omega q}^{(1,2)}(\rho, z)=
\psi_{n\Omega q}^{(1,2)}(\rho, z, \downarrow), \\
 \end{array}
\end{equation}
where wave functions $U$ and $D$ denote
  the spin-up ($\sigma$=$\frac{1}{2}$) and spin-down
($\sigma$=$-\frac{1}{2}$) spinor components. Since the time reversal symmetry is
conserved, one can consider  only positive-$\Omega$ values.
In terms of these wave functions,
the particle density
$\rho_q(\mathbf{r})$ and pairing density
$\tilde{\rho}_q(\mathbf{r})$ can be written as:
\begin{equation}
\begin{array}{l}
\begin{array}{ccl}
\displaystyle \rho_q(\mathbf{r})&=&\displaystyle
\sum_{\sigma}\sum_{n\Omega}\psi_{n\Omega
q}^{(2)}(\mathbf{r}\sigma)\psi_{n\Omega
q}^{(2)*}(\mathbf{r}\sigma)\vspace{3pt}\\
&\displaystyle =& \displaystyle
\frac{1}{\pi}\sum_{\Omega=\frac{1}{2}}^{\Omega_{\rm
max}}\sum_{E_n>0}^{E_{\rm max}}[|U_{n\Omega q}^{(2)}(\rho,z)|^2\\
&&~~~~~~~~~~~~~~~~+|D_{n\Omega
q}^{(2)}(\rho,z)|^2],\\
\end{array}
\vspace{6pt}\\
\begin{array}{ccl}
\displaystyle \tilde{\rho}_q(\mathbf{r})&=&-\displaystyle
\sum_{\sigma}\sum_{n\Omega}\psi_{n\Omega
q}^{(2)}(\mathbf{r}\sigma)\psi_{n\Omega
q}^{(1)*}(\mathbf{r}\sigma)\vspace{3pt}\\
&\displaystyle =& -\displaystyle
\frac{1}{\pi}\sum_{\Omega=\frac{1}{2}}^{\Omega_{\rm
max}}\sum_{E_n>0}^{E_{\rm max}}[U_{n\Omega q}^{(2)}(\rho,z)U_{n\Omega
q}^{(1)*}(\rho,z)\\
&&~~~~~~~~~~~~~~~+D_{n\Omega q}^{(2)}(\rho,z)D_{n\Omega
q}^{(1)*}(\rho,z)].\\
\end{array}
\end{array}
\end{equation}
In the above equations, the sums are limited by the quasiparticle
energy cutoff $E_{\rm max}$, defining the effective range for a
zero-range pairing force,  and the angular projection cutoff
$\Omega_{\rm max}$. This way of truncating the continuum space is
different from that  in the spherical {\hfbrad}
code~\cite{bennaceur} which employs a maximum $j_{\rm max}$ cutoff
on the total single-particle angular momentum.
 This implies that
even with the same energy cutoff, {\hfbax} and {\hfbrad} have
different pairing phase spaces. We shall  return to this point later in
Sec.~\ref{benchmarking}.

\begin{table}[htb]
 \caption{ \label{tablebound} Boundary conditions of HFB wave functions at
 $\rho$=0 and  $z$=0  in {\hfbax}. The single-particle states are labeled
 by $\Omega^\pi$ quantum numbers. }
\begin{ruledtabular}
\begin{tabular}{c|c|c}
&$\rho$=0 ~~~~~~~~~~~~& $z$=0~~~~~~~~~~~ \\ \hline
 $\begin{array}{c}\Omega-\frac{1}{2}={\rm even} \\
\pi=+1 \\ \end{array}$&  $\begin{array}{l} \displaystyle
\frac{\partial U}{\partial \rho }|_{\rho=0}=0
~~~~~ \vspace{2pt} \\  D|_{\rho=0}=0\\
\end{array}  $  & $ \begin{array}{l}
 \displaystyle \frac{\partial U}{\partial z}|_{z=0}=0 ~~~~~ \vspace{2pt} \\
D|_{z=0}=0\\
\end{array}   $\\
\hline
$\begin{array}{c}\Omega-\frac{1}{2}={\rm odd} \\
\pi=+1 \\ \end{array}$&  $\begin{array}{l}
\displaystyle \frac{\partial D}{\partial \rho}|_{\rho=0}=0 ~~~~~ \vspace{2pt} \\
 U|_{\rho=0}=0\\
\end{array} $ &$ \begin{array}{l}
\displaystyle \frac{\partial D}{\partial z}|_{z=0}=0 ~~~~~ \vspace{2pt} \\
U|_{z=0}=0\\
\end{array}$\\
\hline
 $\begin{array}{c} \Omega-\frac{1}{2}={\rm even} \\
\pi=-1 \\ \end{array}$&  $\begin{array}{l}
\displaystyle \frac{\partial U}{\partial \rho}|_{\rho=0}=0 ~~~~~ \vspace{2pt} \\
D|_{\rho=0}=0\\
\end{array} $  &  $ \begin{array}{l}
\displaystyle \frac{\partial D}{\partial z}|_{z=0}=0 ~~~~~\vspace{2pt}  \\
U|_{z=0}=0\\
\end{array}$\\
\hline
 $\begin{array}{c}\Omega-\frac{1}{2}={\rm odd} ~~~~~ \vspace{2pt} \\
\pi=-1  \\
\end{array}$& $\begin{array}{l}
\displaystyle \frac{\partial D}{\partial \rho}|_{\rho=0}=0 ~~~~~\vspace{2pt}  \\
\vspace{-1pt}
U|_{\rho=0}=0\\
\end{array} $  & $ \begin{array}{l}
\displaystyle \frac{\partial U}{\partial z}|_{z=0}=0 ~~~~~ \vspace{2pt} \\
\vspace{-1pt}
D|_{z=0}=0\\
\end{array}$ \\
\end{tabular}
\end{ruledtabular}
\end{table}
In the reflection-symmetric version of the code {\hfbax} discussed in this paper,
we assumed the space inversion as a  self-consistent symmetry. Consequently,
the quasiparticle wave functions are eigenstates of the parity operator
$\hat{\mathcal{P}}$:
\begin{equation}
\hat{\mathcal{P}}\Psi_{n, \Omega, q}(\rho, z, \phi)=\pi \Psi_{n,
\Omega, q}(\rho, -z, \phi+\pi),
\end{equation}
i.e., parity  $\pi=\pm$ is a good quantum number. The presence of
conserved parity implies specific boundary conditions at $\rho$=0
and $z$=0 (see Table~\ref{tablebound}). We also apply the box
boundary conditions at the outer box boundaries; namely, the wave
functions are put to zero at the edge of a 2D box $z_{\rm max}$ and
$\rho_{\rm max}$. These boundary conditions are important for  the
construction of derivative operators.

In a given
($\Omega, q$) block, the HFB Hamiltonian in Eq.~(\ref{HFBm})
can be expressed through the mean fields $h$ and $\tilde{h}$
with specified spin projections~\cite{teran}:
\begin{equation}\label{hupdown}
\left (
\begin{array}{cccc}
h_{\uparrow\uparrow}-\lambda&h_{\uparrow\downarrow}&\tilde{h}_{\uparrow\uparrow}&\tilde{h}_{\uparrow\downarrow}\vspace{3pt}\\
h_{\downarrow\uparrow}&h_{\downarrow\downarrow}-\lambda&\tilde{h}_{\downarrow\uparrow}&\tilde{h}_{\downarrow\downarrow}\vspace{3pt}\\
\tilde{h}_{\uparrow\uparrow}&\tilde{h}_{\uparrow\downarrow}&-h_{\uparrow\uparrow}+\lambda&h_{\uparrow\downarrow}\vspace{3pt}\\
\tilde{h}_{\downarrow\uparrow}&\tilde{h}_{\downarrow\downarrow}&h_{\downarrow\uparrow}&-h_{\downarrow\downarrow}+\lambda\\
\end{array}
\right )
\end{equation}
The local Skyrme p-h Hamiltonian   $h$  has the usual form
\cite{dobaczewski84,dobaczewski96}:
\begin{equation}
\displaystyle
h_q(\mathbf{r},\sigma,\sigma')=-{\boldsymbol \nabla}\cdot\frac{\hbar^2}
{2m^{*}}{\boldsymbol \nabla} +U_q-i\mathbf{B}_q\cdot({\boldsymbol \nabla}
\times {\boldsymbol\sigma})
\end{equation}
where $m^{*}$ is the effective mass,  $U_q$ is the central p-h
mean-field potential including the Coulomb term for protons,  and
 the spin-orbit
potential with
\begin{equation}
\mathbf{B}_q=\frac{1}{2}W_q({\boldsymbol \nabla}\rho(\mathbf{r})+
{\boldsymbol \nabla}\rho_q(\mathbf{r})),
\end{equation}
where $W_q$ is the  spin-orbit coupling strength.

The pairing Hamiltonian $\tilde{h}$ corresponding to  the zero-range
density-dependent $\delta$
 interaction can be written as
\begin{equation}\label{deltapai}
\displaystyle
\tilde{h}_q(\mathbf{r},\sigma,\sigma')=V_0^q\tilde{\rho}_q(\mathbf{r})F(\mathbf{r})\delta_{\sigma\sigma'}
\end{equation}
where $V_0^q<0$ is the pairing strength, and the pairing form factor
 $F(\mathbf{r})$ depends on
the form of pairing Hamiltonian~\cite{dobaczewski02}:
\begin{equation}
F(\mathbf{r})=\left\{
\begin{array}{lcc}
1& \rm -~volume& \rm  pairing\\
\displaystyle 1-\frac{\rho(\mathbf{r})}{\rho_0}& \rm  -~surface&\rm pairing \\
\displaystyle 1-\frac{\rho(\mathbf{r})}{2\rho_0} &\rm  -~mixed &\rm  pairing \\
\end{array}
\right.
\end{equation}
where $\rho_0$=0.16\,fm$^{-3}$. The volume pairing interaction  acts
primarily inside the nuclear volume while the surface pairing
generates pairing fields peaked around or outside  the  nuclear
surface. As discussed in Ref.~\cite{doba01}, different forms of
$F(\mathbf{r})$ can result in notable differences of pairing fields
in diffused drip-line nuclei.

\subsection{B-spline technique in {\hfbax}}

The lattice representation of wave functions and the HFB Hamiltonian
used in this work closely follows that of {\hfbvan} described in
Ref.~\cite{blazkiewicz}. In {\hfbax}, the wave functions are
discretized on  a 2D grid  ($r_{\alpha}$, $z_{\beta}$) with the
$M$-order B-splines:
\begin{equation}
\psi_{n\Omega^\pi q}^{(1, 2)}(\rho_{\alpha}, z_{\beta})=\displaystyle
\sum_{i,j}B^M_i(\rho_{\alpha})B^M_j(z_{\beta})C_{n\Omega^\pi q}^{ij(1,2)},
\end{equation}
where $C^{ij}$ is the matrix of expansion coefficients;
$\alpha=1,\ldots,N_{\rho}$ and $\beta=1,\ldots,N_z$. The four
components of  HFB wave functions (\ref{wf4}) are thus represented
in a matrix form. The derivative operators are constructed using the
Galerkin method. In the B-spline representation, the HFB Hamiltonian
acts on 2D wave functions like a tensor, i.e.,
\begin{equation}
{h}^{\alpha\beta}_{\gamma\delta}\psi(\rho_{\alpha},
z_{\beta})=\psi'(\rho_{\gamma}, z_{\delta}).
\end{equation}
The HFB equation is solved by mapping the Hamiltonian tensor into a
matrix, which is then diagonalized. The associated wave functions
$\langle \Psi^{*}|$   and $|\Psi\rangle $ are mapped into the bra
and ket vectors, respectively:
\begin{equation}\label{braket}
\langle
\Psi^{*}_{\gamma\delta}|\mathcal{H}_{\gamma\delta}^{\alpha\beta}|\Psi_{\alpha\beta}\rangle
=E.
\end{equation}

In the following,  we  give some details pertaining
to the Hamiltonian mapping, because
the mapping rule is different with respect to the 16 individual
blocks in Eq.~(\ref{hupdown}). For the diagonal blocks, it is
straightforward  to map a
tensor into a matrix ($k$, $k'$):
\begin{widetext}
\begin{equation}
\begin{array}{l}
\begin{array}{cccc}
&~~h(\uparrow\uparrow):  &\left\{
\begin{array}{c}
k=(\beta-1)N_{\rho}+\alpha \\
k'=(\delta-1)N_{\rho}+\gamma \\
\end{array}
\right.
\end{array} \vspace{2pt}
\\
\begin{array}{lllc}
&~~h(\downarrow\downarrow): &\left\{
\begin{array}{l}
k=(\beta-1)N_{\rho}+\alpha+N_\rho N_z \\
k'=(\delta-1)N_{\rho}+\gamma+N_\rho N_z  \\
\end{array}
\right.
\end{array} \vspace{2pt}
\\
\begin{array}{cccc}
&-h(\uparrow\uparrow):  &\left\{
\begin{array}{c}
k=(\beta-1)N_{\rho}+\alpha+2\times N_\rho  N_z  \\
k'=(\delta-1)N_{\rho}+\gamma+2\times N_\rho N_z  \\
\end{array}
\right.
\end{array} \vspace{2pt}
\\
\begin{array}{cccc}
&-h(\downarrow\downarrow): &\left\{
\begin{array}{c}
k=(\beta-1)N_{\rho}+\alpha+3\times N_\rho N_z  \\
k'=(\delta-1)N_{\rho}+\gamma+3\times N_\rho N_z  \\
\end{array}
\right.
\end{array}
\\
\end{array}
\end{equation}

 Following the
same rule, the bra  and ket  vectors are mapped into
vectors with indexes $k$ and $k'$, respectively.
It is more complicated to map the off-diagonal blocks
of (\ref{hupdown}).
For the four ket  vectors,  the mapped index $k$ should correspond
to different columns of the Hamiltonian blocks. For example,
the index of the first ket component $U^{(1)}$ should
correspond to the upper index of the first column of (\ref{hupdown}):
\begin{equation}
\begin{array}{cccc}
(U^{(1)*}&D^{(1)*}&U^{(2)*}&D^{(2)*})\\
\end{array}
\left(
\begin{array}{|c|ccc}
\overline{h^{\alpha\beta}(\uparrow\uparrow)-\lambda}&h(\uparrow\downarrow)&\tilde{h}(\uparrow\uparrow)&\tilde{h}(\uparrow\downarrow)\vspace{0pt}\\
h^{\alpha\beta}(\downarrow\uparrow)&h(\downarrow\downarrow)-\lambda&\tilde{h}(\downarrow\uparrow)&\tilde{h}(\downarrow\downarrow)\vspace{0pt}\\
\tilde{h}^{\alpha\beta}(\uparrow\uparrow)&\tilde{h}(\uparrow\downarrow)&-h(\uparrow\uparrow)+\lambda&-h(\uparrow\downarrow)\vspace{0pt}\\
\underline{~~~\tilde{h}^{\alpha\beta}(\downarrow\uparrow)~~~}&\tilde{h}(\downarrow\downarrow)&-h(\downarrow\uparrow)&-h(\downarrow\downarrow)+\lambda\\
\end{array}
\right) \left(
\begin{array}{l}
\hline
U^{(1)}(\rho_{\alpha}z_{\beta})\\
\hline
D^{(1)}\\
U^{(2)}\\
D^{(2)}\\
\end{array}
\right).
\end{equation}
For the four bra vectors,  the mapped index $k'$ corresponds
to different rows of the Hamiltonian. For example,
 the index of the first bra  component $U^{(1)*}$
corresponds to the lower index of the first row of (\ref{hupdown}):
\begin{equation}
\begin{array}{cccc}
( \underline{\overline{|U^{(1)*}(\rho_{\gamma}z_{\delta})}}|&D^{(1)*}&U^{(2)*}&D^{(2)*})\\
\end{array}
\left(
\begin{array}{cccc}
\hline
h_{\gamma\delta}(\uparrow\uparrow)-\lambda&h_{\gamma\delta}(\uparrow\downarrow)&
\tilde{h}_{\gamma\delta}(\uparrow\downarrow)&\tilde{h}_{\gamma\delta}(\downarrow\downarrow)\vspace{0pt}\\
\hline
h(\downarrow\uparrow)&h(\downarrow\downarrow)-\lambda&\tilde{h}(\downarrow\uparrow)&\tilde{h}(\downarrow\downarrow)\vspace{0pt}\\
\tilde{h}(\uparrow\uparrow)&\tilde{h}(\uparrow\downarrow)&-h(\uparrow\uparrow)+\lambda&-h(\uparrow\downarrow)\vspace{0pt}\\
\tilde{h}(\downarrow\uparrow)&\tilde{h}(\downarrow\downarrow)&-h(\downarrow\uparrow)&-h(\downarrow\downarrow)+\lambda\\
\end{array}
\right)
 \left(
\begin{array}{c}
U^{(1)}\\
D^{(1)}\\
U^{(2)}\\
D^{(2)}\\
\end{array}
\right).
\end{equation}
Following this scheme, the HFB Hamiltonian, represented by 16 tensor
blocks,
 is mapped into a matrix of
the rank $4\times N_{\rho}N_z$. The resulting
 HFB Hamiltonian matrix is diagonalized using {\sc lapack}
routines~\cite{lapack}.
\end{widetext}

The Coulomb potential is obtained by directly integrating the
Poisson equation,
\begin{equation}
\nabla^2\phi(\rho,z)=-4\pi e^2 \rho_p(\rho,z),
\end{equation}
where $\phi$ is the Coulomb potential and  $\rho_p$ is the proton
density. The Poisson equation, discretized  on a B-spline grid,  can
be written in a matrix form. The boundary conditions  at large
distances are given by the multipole expansion of the Coulomb
potential~\cite{blazkiewicz}. The gradient of the Coulomb potential
at $z$=0 or $\rho$=0 is set to be zero because of the symmetries
imposed. These boundary conditions are incorporated in the lattice
representation of the Laplace operator.

\subsection{Numerical speedup}

The size of the HFB Hamiltonian matrix  depends on the box size $R$,
the largest distance between neighboring mesh points in the grid $h$
(the B-spline grid is not uniform), and the order of B-splines $M$.
Consequently, for large and refined grids, the diagonalization time
becomes a  bottleneck. For example, for $^{120}$Sn with a box of
19.2\,fm, a maximum grid size of $h$=0.6 fm, and $M$=13, the
Hamiltonian matrix is about 0.45 GB of storage, and one
diagonalization takes about 30 CPU minutes.

Since the  diagonalization of HFB matrices corresponding to
different $\Omega^\pi$ blocks can be  done independently,
this part of {\hfbax} can be parallelized using the
Message Passing Interface. For $^{120}$Sn with
$\Omega_{\rm max}$=33/2 cutoff, about 70 processors are utilized.
 The precision of the derivative operators is
crucial for the accuracy of the HFB eigenstates. By distributing
diagonalization over many processors, one can perform calculations
with  larger boxes, denser grids, and higher-order B-splines. In
addition, by taking advantage of
 the reflection symmetry,  $N_z$ can be reduced; hence  the rank of the Hamiltonian matrix
is scaled down by  a factor slightly
 less than 2. Consequently, the
diagonalization process, which takes most of the execution time, can
be speeded up by a factor  greater than 8.

For  the diagonalization, we employ the {\sc lapack} {\sc dgeev}
routine \cite{lapack}. This routine diagonalizes a non-symmetric
matrix using a QR algorithm. However, due to energy cutoff in HFB,
it is not necessary to compute all eigenvectors. For this reason, we
modified {\sc dgeev} so that  it yields eigenvectors only within the
required energy  window. By this way, the diagonalization time for
$^{120}$Sn is further reduced by one-third.

To optimize the  convergence of HFB iterations, we take the HF-BCS
densities to warm-start  the self-consistent process. Furthermore,
instead of the commonly used linear mixing, we have implemented the
modified Broyden mixing, quasi-Newton algorithm to solve large sets
of non-linear equations, for accelerating the convergence
rate~\cite{broyden}. The iterative convergence can be estimated by
the input and output difference at the $m$-th iteration,
\begin{equation}\label{broydenm}
\Fvec^{(m)}=\Vvec_{out}^{(m)}-\Vvec_{in}^{(m)},
\end{equation}
where $\Vvec$ is an $N$-dimensional vector containing unknowns, and
self-consistency condition requires that the solution $\Vvec^{*}$ be
a fixed-point of the  iteration: $\Ivec(\Vvec^{*}) = \Vvec^{*}$. For
the linear mixing, the  input at iteration $m$+1 is given as
\begin{equation}\label{linearm}
\Vvec_{in}^{(m+1)}=\Vvec_{in}^{(m)}+\alpha \Fvec^{(m)},
\end{equation}
where $\alpha$ is a constant between 0 and 1. In contrast, to
estimate the next  step, the modified Broyden mixing utilizes
information obtained in previous  $M_B$ iterations. Recent
implementations of this technique to the HFB problem have
demonstrated that the sufficient convergence  can generally be
obtained within 20-30 iterations~\cite{baran}. In {\hfbax} the
vector $\Vvec$ consists of local densities and their derivatives at
lattice mesh points:
\begin{equation}
\Vvec \equiv \left\{
\rho_{q}, \tau_{q}, {\boldsymbol\nabla}\cdot\mathbf{J}_{q},
\tilde{\rho}_{q},\nabla^2\rho_{q}, \nabla_\rho\rho_{q},
\nabla_z\rho_{q}
 \right \}.
\end{equation}
The dimension of $\Vvec$ is thus 14$\times N_\rho \times  N_z$.

\begin{figure}[htb]
\centerline{\includegraphics[trim=0cm 0cm 0cm 0cm,width=0.45\textwidth,clip]{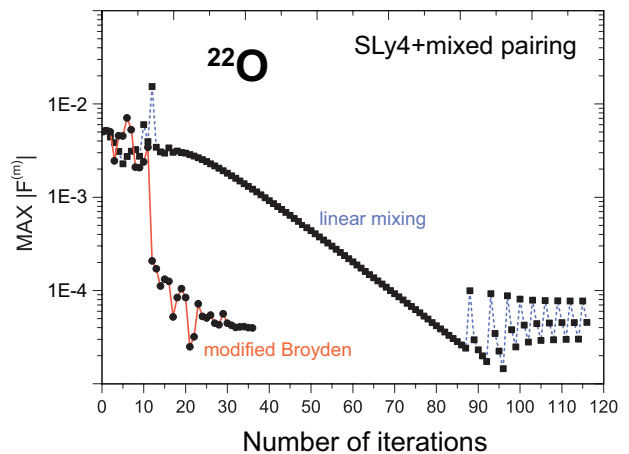}}
\caption{\label{byd} (Color online) Comparison between linear mixing
(squares) and modified Broyden's method (circles) in {\hfbax} for
$^{22}$O. The largest element of $|\Fvec^{(m)}|$ is shown as a
function of the number of iterations  $m$. See text for details.}
\end{figure}

To demonstrate the performance of the method, we consider  $^{22}$O
in a 2D square box of $R$=12\,fm, $h$=0.6\,fm and $M$=11. The
calculations were carried out on  a Cray XT3 supercomputer at ORNL
having 2.6 GHz AMD processors. Without reflection symmetry, one
diagonalization takes about 15 minutes of CPU time. With reflection
symmetry imposed, one diagonalization needs about 100 seconds with
the modified {\sc dgeev}. The calculations with Broyden's mixing
with $M_B$=7 are displayed in Fig.~\ref{byd}. The actual variation
of binding energy is within 0.1 keV after 30 iterations. This is to
be compared with linear mixing  with $\alpha$=0.6 which requires
over 80 iterations to reach similar convergence. Both calculations
show precision limitations due to the   numerical noise inherent to
the mesh assumed. As in examples discussed in Ref.~\cite{baran},
Broyden's method implemented in {\hfbax}
  provides impressive performance
improvements of  HFB iterations. The numerical speedup is particularly
helpful for heavy nuclei and for constrained calculations, which usually require
 many  self-consistent iterations.
%
% trim=l b r t

\section{Benchmarking of {\hfbax} and typical applications}\label{benchmarking}

This section contains results of {\hfbax} test results.
First, the absolute accuracy of the one-body HF solver is
 tested
using the adaptive multi-resolution method. Thereafter follows a
series of calculations in which {\hfbax} is benchmarked against the
spherical coordinate space code {\hfbrad} and the axial code
{\hfbtho}. In all those realistic tests, the Skyrme functional SLy4
\cite{sly4}   was used in the p-h channel, augmented by  different
density-dependent delta functionals (\ref{deltapai}) in the p-p
channel.

\subsection{Two-center potential:  comparison with HO and
wavelet expansions}\label{wavelets}

The accuracy of the {\hfbax} calculations in the p-h channel has been tested using
an  adaptive  multiwavelet basis. To this end,
we employed
the {\sc madness} framework \cite{madness}. The details regarding
our particular realization of the
wavelet basis expansion can be found in Refs.~\cite{wavelet,fann}.

\begin{figure}[htb]
\centerline{\includegraphics[trim=0cm 0cm 0cm 0cm,width=0.4\textwidth,clip]{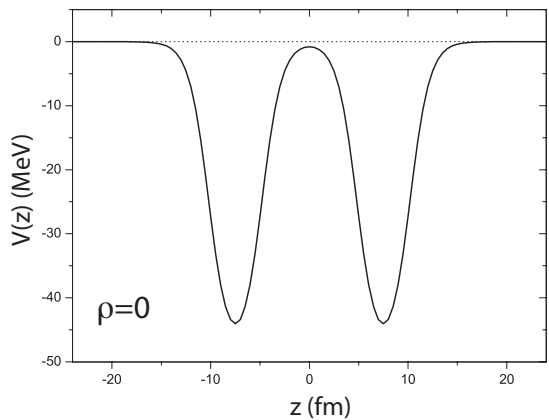}}
\caption{ \label{fig1:twocosh} The two-center inverted-cosh
potential of Eq.~(\ref{tcosh}) as a function of $z$ at $\rho$=0. The
two centers are 15\,fm apart ($\zeta$=7.5\,fm) and $V_0$=$-$50\,MeV,
$R_0$=2\,fm,  and $a$=1\,fm.}
\end{figure}

As a test case, we choose
an axial  two-center inverted-cosh potential:
\be\label{tcosh}
V(\rho,z)=V_0 \left[f(\rho,z+\zeta) + f(\rho,z-\zeta)\right],
\ee
where the inverted-cosh form factor is:
\be
f(\rho,z) = \frac{1}{1+e^{-R_0/a}\cosh(\sqrt{\rho^2+z^2}/a)},
\ee
and $V_0$, $R_0$, and $a$ are the potential's depth, radius, and
diffuseness, respectively, and $2\zeta$ is the distance between the
two centers. A cross section of the potential used in
our test calculations
is shown in
Fig.~\ref{fig1:twocosh} along the $z$-axis  at $\rho$=0. It is seen that
 the two potential's centers
are well separated; hence the ability to predict a small parity splitting
between the  eigenstate doublets provides a stringent test for
the B-spline Schr\"odinger equation solver of {\hfbax}.
%
% trim=l b r t

In addition to  potential (\ref{tcosh}) we also considered the
spin-orbit term in the usual Thomas form:
\begin{equation}
\displaystyle
V_{SO}(\rho,z)=-i\lambda_0\left(\frac{\hbar}{2mc}\right)^2
{\boldsymbol \nabla}
V(\rho,z)\cdot({\boldsymbol \sigma}\times {\boldsymbol \nabla}),
\end{equation}
where $\lambda_0$=5.0 and the numerical values of fundamental  constants
 were taken as
$\hbar^2/2m$=20.721246 fm$^2$, $\hbar c$=197.32696 MeV fm,
and $mc^2$=939.56535 MeV.
The corresponding one-body Schr\"{o}dinger equation reads:
\begin{equation}
\left [ \displaystyle -\frac{\hbar^2}{2m}\nabla^2+V(\rho, z)+V_{SO}(\rho,
z)\right ]\varphi(\rho, z, \phi)=E\varphi(\rho, z, \phi),
\end{equation}
where $\varphi(\rho, z, \phi)$ is a two-component spinor wave function.

\begin{table}[htb]
 \caption{\label{tabho1}
 Ten lowest eigenvalues of the
 two-center potential (\ref{tcosh})  obtained using the
  HO, B-spline,  and wavelet expansions. All energies are in MeV.
  For more details see text.}
\begin{ruledtabular}
\begin{tabular}{cccccc}
State No.& $\Omega^{\pi}$ & HO & HO & B-spline  & Wavelets\\
&& $N_{sh}$=20 & $N_{sh}$=30 & $h$=0.6& \\
 \hline
1&1/2$^{+}$&$-$22.23916&$-$22.24008 &$-$22.24011&$-$22.24011\\
2&1/2$^{-}$&$-$22.23816&$-$22.23995&$-$22.23998&$-$22.23998\\
3&1/2$^{+}$&$-$9.21514&$-$9.22047&$-$9.22050&$-$9.22050\\
4&3/2$^{-}$&$-$9.20359& $-$9.21256&$-$9.21260&$-$9.21260\\
5&1/2$^{-}$&$-$9.20359& $-$9.21256&$-$9.21260&$-$9.21260\\
6&3/2$^{+}$&$-$9.20589&$-$9.21126&$-$9.21129&$-$9.21129\\
7&1/2$^{+}$&$-$9.20589&$-$9.21126&$-$9.21129&$-$9.21129\\
8&1/2$^{-}$&$-$9.19743& $-$9.20590&$-$9.20595&$-$9.20595\\
9&1/2$^{+}$&$-$1.70724&$-$1.72402&$-$1.72503&$-$1.72514\\
10&1/2$^{-}$&$-$1.49218&$-$1.52486&$-$1.52672&$-$1.52690\\
\end{tabular}
\end{ruledtabular}
\end{table}
Table~\ref{tabho1}  displays the lowest eigenvalues of the
 two-center potential (\ref{tcosh})  obtained in three
expansion methods. In the HO expansion calculation, we took
$N_{sh}$=20 and 30 shells of the spherical oscillator with
$\hbar\omega_0$=5.125\,MeV (as it turned out, the use of a stretched
basis was not particularly advantageous). The size of the
$\Omega^\pi$=${\frac{1}{2}}^+$ Hamiltonian block is 121 and 256 for
$N_{sh}$=20 and 30, respectively, i.e., the matrix size is more than
doubled in the latter case. In the  {\hfbax} calculation with
$M$=13, we used a square box of $R$=25.2\,fm and $h$=0.6\,fm. (The
values of $N_{sh}$=20 in HO and $h$=0.6\,fm in {\hfbax} are typical
for realistic Skyrme-HFB calculations.) In the wavelet variant
\cite{wavelet,fann}, the absolute accuracy was assumed to be
 10$^{-5}$.

It is seen that the accuracy of B-spline expansion
 is excellent, both for the absolute energies and for the
 parity splitting.  The HO basis with $N_{sh}$=20
performs rather poorly, especially for parity splitting and for the
energies of the highest (halo) states. That is, of course, to be
expected for a two-center potential expanded in a one-center basis.
It is only at $N_{sh}$=30, not practical in large-scale DFT
calculations, that a good agreement with wavelets and {\hfbax} is
obtained. In all variants, there is a  perfect degeneracy of
$p_{3/2}$ doublets with $\Omega$=$\frac{1}{2}$  and
$\Omega$=$\frac{3}{2}$  .

\begin{table}[htb]
 \caption{\label{tabho2} Similar to Table~\ref{tabho1}
except for the two-center double-cosh
potential with the spin-orbit term.}
\begin{ruledtabular}
\begin{tabular}{cccccc}
State No.& $\Omega^{\pi}$& HO & HO & B-spline  & Wavelets \\
&& $N_{sh}$=20 & $N_{sh}$=30 & $h$=0.6 & \\
\hline
1&1/2$^{+}$&$-$22.23916&$-$22.24008&$-$22.24011&$-$22.24011\\
2&1/2$^{-}$&$-$22.23816&$-$22.23995&$-$22.23998&$-$22.23998\\
3&1/2$^{+}$&$-$9.43145&$-$9.43659&$-$9.43663&$-$9.43662\\
4&3/2$^{-}$&$-$9.42314&$-$9.43199&$-$9.43203&$-$9.43202\\
5&3/2$^{+}$&$-$9.42561&$-$9.43078&$-$9.43081&$-$9.43080\\
6&1/2$^{-}$&$-$9.41931&$-$9.42783&$-$9.42788&$-$9.42788\\
7&1/2$^{+}$&$-$8.77250&$-$8.77825&$-$8.77828&$-$8.77828\\
8&1/2$^{-}$&$-$8.76475&$-$8.77380&$-$8.77384&$-$8.77383\\
9&1/2$^{+}$&$-$1.70727&$-$1.72405&$-$1.72506&$-$1.72516\\
10&1/2$^{-}$&$-$1.49222&$-$1.52490&$-$1.52675&$-$1.52693\\
\end{tabular}
\end{ruledtabular}
\end{table}
The results with the inclusion of the spin-orbit term, which lifts
the degeneracy between $\Omega$=$\frac{1}{2}$  and $\frac{3}{2}$
levels, are given in
Table~\ref{tabho2}. It is seen that the general excellent agreement
between B-spline and wavelet variants holds, and that HO with
$N_{sh}$=20 performs rather poorly, especially for the halo state.
%
% trim=l b r t
\begin{figure}[htb]
\centerline{\includegraphics[trim=0cm 0cm 0cm 0cm,width=0.4\textwidth,clip]{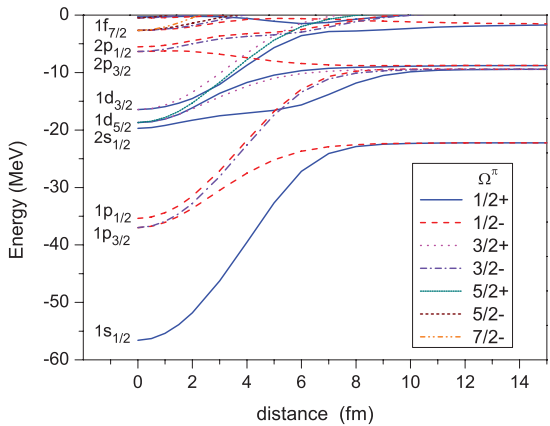}}
\caption{\label{twocoshsp}(Color online) Eigenvalues of a two-center
inverted-cosh potential with the spin-orbit term calculated in
{\hfbax} as functions of the distance between two centers $2\zeta$.}
\end{figure}
Finally, the single-particle
spectrum of a two-center potential is illustrated in
Fig.~\ref{twocoshsp} as a function of the inter-center distance.
A transition to a dimer-like spectrum is clearly seen at
 distances greater than 6\,fm. Our tests nicely illustrate the
ability of the B-spline technique (thus: {\hfbax}) to handle a
two-center problem encountered, e.g., in fission or fusion
\cite{models}.

\subsection{Spherical limit: comparison with {\hfbrad} and {\hfbtho}}
\label{stablesph}

The performance  of {\hfbax}  at the spherical limit
 can be assessed by comparing it
against the accurate 1D radial coordinate code {\hfbrad}, based on a
direct integration of the system of coupled radial differential
equations \cite{dobaczewski84,bennaceur}. The  tests have been
carried out for the nucleus $^{120}$Sn, which is often used in
benchmarking HFB solvers \cite{dobaczewski04, blazkiewicz} in the
limit of spherical shape and nonzero neutron pairing.

The precision of HFB calculations in coordinate space is primarily
determined by the size of the mesh  used. We calculated  $^{120}$Sn
with the fixed-box size ($R$=19.2 fm) but with different mesh steps
and B-spline orders. In our  calculations,  we took the volume delta
pairing interaction with the pairing strength  $V_0$=$-$187.05 MeV
fm$^{3}$ adjusted to the experimental neutron pairing gap of
 $\Delta_n$=1.245 MeV.
For the pairing space, we adopted the commonly used equivalent
energy cutoff of 60 MeV \cite{dobaczewski04}. As both codes are
written in different geometry, the quasiparticle continuum is
discretized differently in {\hfbrad} and {\hfbax}. In {\hfbrad}, all
partial waves with $j$$\le$$j_{\rm max}$=33/2 were considered, while
in {\hfbax} we imposed  a cutoff on $j_z$: $\Omega_{\rm max}$=33/2.
 For the sake of comparison,  the pairing regularization option in
{\hfbrad} has been  turned off.
  Also, we adopted  the same fundamental constants as  in
Ref~\cite{dobaczewski04}:
$\hbar^2/2m=20.73553$ MeV and $e^2$=1.439978 MeV fm.

\begin{table}[htb]
 \caption{\label{sntab1} Results of spherical
 HFB+SLy4 calculations with volume pairing for  $^{120}$Sn using
 {\hfbax} with
 different mesh size $h$ and
  B-spline order $M$. The results of the precision radial code {\hfbrad} are
  shown for comparison. The static  proton pairing is zero.
  All energies are in MeV;
  $h$ is in fm.}
\begin{ruledtabular}
\begin{tabular}{cccccc}
$E$ & $h$=0.64  & $h$=0.6 & $h$=0.6 & $h$=0.15   \\
    & $M$=11    & $M$=11  & $M$=13  & {\hfbrad}  \\
\hline
~~$E_{tot}$&-1018.304&$-$1018.271&$-$1018.356&$-$1018.362\\
~~$E_{C}$&347.636&347.642&347.577&347.558\\
~~$E_{kin}^p$&831.520&831.512&831.534&831.520\\
~~$E_{kin}^n$&1339.516&1339.553&1339.562&1339.598\\
~~$E_{pair}^n$&$-$10.253&$-$10.253&$-$10.253&$-$10.278\\
~~$\tilde{E}_{kin}^n$ & 1329.263 & 1329.300 & 1329.309 & 1329.320\\
~~$\Delta_n$&1.2451&1.2451&1.2451&1.2450\\
~~$\lambda_n$&$-$7.9950&$-$7.9950&$-$7.9950&$-$7.9953\\
\end{tabular}
\end{ruledtabular}
\end{table}
Table~\ref{sntab1} displays various contributions to the binding energy $E_{tot}$
of $^{120}$Sn, i.e.,   kinetic
energy $E_{kin}$ for protons and neutrons, Coulomb energy $E_{C}$, neutron pairing
energy $E_{pair}^n$,  neutron pairing gap $\Delta_n$ and Fermi
level  $\lambda_n$, and the sum
\be\label{ereg}
\tilde{E}_{kin}=E_{kin}+E_{pair}
\ee
for  neutrons.
(As discussed in Refs.~\cite{Mar98,Pap99,Bor06}, while
for zero-range  pairing  the individual values of
 $E_{kin}$ and $E_{pair}$
are  divergent with respect
to the  cutoff energy of the pairing window,
 their sum (\ref{ereg}) converges nicely
and  it is less sensitive to the actual treatment of discretized
quasiparticle continuum.) The general agreement between {\hfbax} and
{\hfbrad} is excellent, in particular in the $h$=0.6\,fm, $M$=13
variant, with most quantities agreeing within 10\,keV. As expected,
the largest difference is seen for $E_{kin}^n$ and $E_{pair}^n$ due
to a slightly different character of unbound spectrum in the two
models; however, the sum $\tilde{E}_{kin}$ is well reproduced by
{\hfbax}.

The HFB results
with mixed pairing obtained in {\hfbax}, {\hfbrad}, and
{\hfbtho} are shown in Table~\ref{sntab2}.
\begin{table}[htb]
 \caption{\label{sntab2}
  Similar to Table~\ref{sntab1}
except for  the mixed pairing interaction.
{\hfbax} results are compared with those of
{\hfbrad} and  {\hfbtho} of Ref.~\cite{dobaczewski04}.}
\begin{ruledtabular}
\begin{tabular}{cccc}
$E$ &  $h$=0.6 & $h$=0.1 &   \\
    & $M$=13  & {\hfbrad}  & {\hfbtho} \\
\hline
~~$E_{tot}$&$-$1018.795&$-$1018.791&$-$1018.777\\
~~$E_{C}$&347.442&347.400&347.370\\
~~$E_{kin}^p$&830.856&830.848&830.735\\
~~$E_{kin}^n$&1340.675&1340.668&1340.458\\
~~$E_{pair}^n$&$-$12.491&$-$12.467&$-$12.467\\
~~$\tilde{E}_{kin}^n$ & 1328.184 &  1328.201 & 1327.991 \\
~~$\Delta_n$&1.2448&1.2446&1.2447\\
~~$\lambda_n$&$-$8.0186&$-$8.0181&$-$8.0168\\
\end{tabular}
\end{ruledtabular}
\end{table}
The pairing strength in {\hfbax}  was
taken  as  $V_0$=$-$284.29 MeV fm$^3$,
as compared to $-$284.36 and
$-$284.10 MeV fm$^3$
in {\hfbrad} and {\hfbtho}, respectively~\cite{dobaczewski04}.
In  {\hfbtho} calculations,  25 shells of transformed HO
basis were used.
Again, the agreement between  {\hfbax} and {\hfbrad} is excellent, and
the total binding energies obtained in the three methods agree within\,12 keV.
As expected,
the largest differences are seen for $E_{kin}$.
In particular,
{\hfbtho} underestimates the neutron (proton) kinetic energy  by
 about 200 (100) keV. This deviation
  is partly  due to different
representations of the kinetic energy operator in the coordinate space and
in the transformed oscillator basis, and partly
 due to the different continuum space (see the discussion above).

\subsection{Weak binding regime: comparison with {\hfbrad}}\label{nickl}

Neutron-rich  nuclei are unique laboratories of neutron pairing. In
weakly bound nuclei, pairing fields are affected by the coupling to
the continuum space, and this coupling can significantly modify pair
distributions~\cite{dobaczewski96,antipair,meng,doba01}. In
Sec.~\ref{stablesph} we demonstrated that {\hfbax} performs very
well for a stable spherical nucleus $^{120}$Sn. To evaluate the
performance  of {\hfbax} for weakly bound  nuclei, in this section
we discuss  ground-state properties of even-even $^{84,86,88,90}$Ni
isotopes, which are expected to be weakly
bound~\cite{Ter96,Ter07,stoitsov07,Mar08}.

In our test calculations, we adopted the surface pairing with
strength adjusted to $^{120}$Sn ($V_0$=$-$512.6\,MeV\,fm$^3$).
Results of {\hfbrad} and {\hfbax} calculations are listed in
Table~\ref{nitab1}. The nucleus $^{90}$Ni is a drip-line system and
its stability is strongly influenced by pairing. Indeed, it is only
bound with surface pairing; with  volume and mixed pairing,
$^{90}$Ni is calculated to have a positive neutron chemical
potential (see Ref.~\cite{doba01} for more discussion concerning
this point).
%
% trim=l b r t
\begin{figure}[htb]
\centerline{\includegraphics[trim=0cm 0cm 0cm 0cm,width=0.4\textwidth,clip]{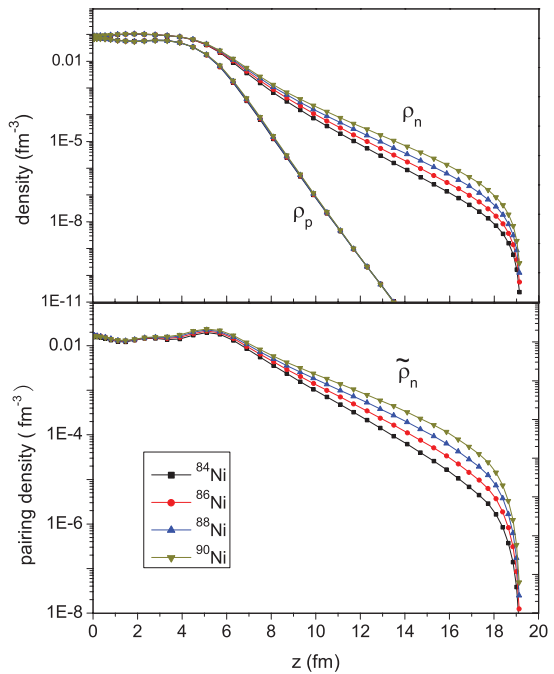}}
\caption{\label{ni} (Color online)Proton  (top) and  neutron pairing
(bottom) densities calculated in {\hfbax} for $^{84,86,88,90}$Ni.
Proton pairing is zero.}
\end{figure}
The local particle and pairing densities of drip-line even-even
Ni isotopes are shown in Fig.~\ref{ni}. A gradual increase of neutron skin
as approaching $^{90}$Ni is clearly seen. The proton density, on the other
hand, is only weakly affected by the outermost neutrons.

\begin{table*}
 \caption{\label{nitab1}Comparison between {\hfbax} and {\hfbrad} with SLy4 p-h
 functional and surface pairing
 for drip-line  nuclei $^{84,86,88,90}$Ni. The same box $R$=19.2\,fm
 was used  in both cases.
 The angular momentum cutoff was taken at $j_{\rm
 max}$=33/2 in {\hfbrad} and $\Omega_{\rm max}$=33/2 in {\hfbax}.
 All energies are in MeV.}
\begin{ruledtabular}
\begin{tabular}{lccccccccc}
~& \multicolumn{2}{c}{$^{84}$Ni}& \multicolumn{2}{c}{$^{86}$Ni}&\multicolumn{2}{c}{$^{88}$Ni}&\multicolumn{2}{c}{$^{90}$Ni}&\\
 $E$ & {\hfbrad}& {\hfbax}&{\hfbrad}& {\hfbax}&{\hfbrad}&{\hfbax}&{\hfbrad}&{\hfbax}\\
 \hline
~$E_{tot}$&$-$654.919&$-$654.899&$-$656.915&$-$656.955&$-$658.215&$-$658.193&$-$658.877&$-$658.856\\
~$E_{C}$&122.797&122.806&122.215&122.228&121.621&121.640&121.018&121.056\\
~$E_{kin}^p$&430.468&430.460&426.311&426.330&422.152&422.206&418.027&418.204\\
~$E_{kin}^n$&1084.511&1084.577&1116.835&1116.782&1148.387&1148.179&1179.697&1178.956\\
~$E_{pair}^n$&$-$30.892&$-$30.890&$-$36.733&$-$37.080&$-$43.179&$-$43.727&$-$49.926&$-$50.807\\
~$\tilde{E}_{kin}^n$ & 1053.619 & 1053.687 & 1080.102 & 1079.702 & 1105.208 & 1104.452 & 1129.771 & 1128.149 \\
~$\Delta_n$&1.485&1.486&1.613&1.617&1.742&1.746&1.862&1.864\\
~$\lambda_n$&$-$1.455&$-$1.454&$-$1.062&$-$1.068&$-$0.709&$-$0.718&$-$0.399&$-$0.417\\
\end{tabular}
\end{ruledtabular}
\end{table*}
The systematic comparison between {\hfbax}  and {\hfbrad}
is given in Table~\ref{nitab1}. For the binding energy, the
agreement is very good, including the borderline system
$^{90}$Ni. The neutron pairing energy increases as one
approaches the neutron drip line. This is consistent with the
systematic behavior of  pairing densities shown in Fig.~\ref{ni}.

For  $^{84}$Ni, {\hfbax} and {\hfbrad}  yield similar
 pairing properties
and kinetic energies. However, with increasing neutron number, the
difference between the values of $\tilde{E}_{kin}^n$ obtained in the
two models gradually grows, reaching over 1.6\,MeV in $^{90}$Ni. At
the same time, the difference between $E_{kin}^p$ values is smaller
by an order of magnitude. The systematic difference between {\hfbax}
and {\hfbrad}  when approaching the neutron drip line can be traced
back to different  angular momentum truncations, i.e., pairing phase
space structure. In {\hfbrad}, the s.p. angular momentum cutoff is
$j_{\rm max}$=33/2, while in the {\hfbax}, the cutoff is done in
terms of the s.p. angular momentum projection and it is $\Omega_{\rm
max}$=33/2. Consequently, in {\hfbax} contributions from high-$j$,
low-$\Omega$ continuum states, absent in {\hfbrad}, are present. In
Sec.~\ref{stablesph}, we have shown that for the well-bound  nucleus
$^{120}$Sn, this difference in the continuum phase space is
insignificant.  However, for nuclei close to the drip line, where
the contribution from unbound states is far more important, the
situation is very different.

\begin{table}
 \caption{\label{nitab3}
 Similar to Table~\ref{nitab1}
except for  $^{90}$Ni with
the angular momentum cutoff  $j_{\rm max}$=49/2 in {\hfbrad} and
$\Omega_{\rm max}$=49/2 in {\hfbax}.
 All energies are in MeV.}
\begin{ruledtabular}
\begin{tabular}{lcc}
 & {\hfbrad} & {\hfbax} ~~\\
\hline
~~$E_{tot}$&$-$658.911 & $-$658.881\\
~~$E_{C}$& 121.038&121.060 \\
~~$E_{kin}^p$&418.165&418.233\\
~~$E_{kin}^n$&1179.073&1178.843 \\
~~$E_{pair}^n$& $-$50.326&$-$50.892 \\
~~$\tilde{E}_{kin}^n$ & 1128.747 & 1127.951 \\
~~$\Delta_n$& 1.860&1.864\\
~~$\lambda_n$&$-$0.410 &$-$0.420\\
\end{tabular}
\end{ruledtabular}
\end{table}
In order to quantify this point, we
performed calculations for
$^{90}$Ni with $j_{\rm max}$=49/2 in
{\hfbrad} and $\Omega_{\rm max}$=49/2 in {\hfbax}.
The results are displayed in Table~\ref{nitab3}.
 The variations in the proton kinetic energy between
 various variants of calculations are small, suggesting
 that the kinetic energy operator is well represented
 by {\hfbrad} and {\hfbax} with the grids assumed.
 Also, the binding energy  changes little, $\sim$30\,keV,
when  the $j$- or $\Omega$-cutoff is increased. In the larger
$j$-window,  $\tilde{E}_{kin}^n$ in  {\hfbrad} is reduced by about
1\,MeV; the corresponding change in {\hfbax} is much smaller,
$\sim$200 keV. This result indicates that the high-$j$ continuum
contributions play an important role in the structure of  $^{90}$Ni.
A need for an appreciable angular momentum cutoff in the description
of weakly bound nuclei, especially for surface-like pairing
interactions, has been pointed  out in  Ref.~\cite{Rotival}.

\subsection{Deformed, weakly bound case: comparison with {\hfbtho}}

One of the objectives  of {\hfbax} is  to precisely solve HFB
equations for axially deformed nuclei, in particular at very large
deformations and/or at the limit of weak binding. In this context,
the neutron-rich Zr isotopes with A$\thicksim$110 are very useful
testing grounds, as they are known to have very large prolate
deformations \cite{blazkiewicz}. In this section,  we compare axial
{\hfbax} and {\hfbtho} calculations for  the nuclei $^{102,110}$Zr
which exhibit
 deformed neutron skin.

\begin{table}[htb]
 \caption{\label{zrtab1}Results of deformed {\hfbax} and {\hfbtho} HFB+SLy4
calculations for $^{102}$Zr and $^{110}$Zr with  mixed pairing. All
energies are in MeV. The quadrupole moments  are in  fm$^2$.}
\begin{ruledtabular}
\begin{tabular}{lcccc}
& \multicolumn{2}{c}{$^{102}$Zr}& \multicolumn{2}{c}{$^{110}$Zr}~~~~~~\\
  & {\hfbax}  & {\hfbtho} & {\hfbax}  & {\hfbtho}\\
\hline
~~~$E_{tot}$&$-$859.649&$-$859.540&$-$893.983&$-$893.840~~~\\
~~~$E_{C}$& 231.149&231.084&226.758&226.712~~~\\
~~~$E_{kin}^p$& 651.309&651.099&632.115&631.882~~~\\
~~~$E_{kin}^n$& 1202.050&1201.990&1368.206&1368.201~~~\\
~~~$E_{pair}^n$&$-$3.261&$-$3.535&$-$3.200&$-$3.323~~~\\
~~~$\tilde{E}_{kin}^n$&1198.789& 1198.455& 1365.006 & 1364.878~~~\\
~~~$\Delta_n$&0.672&0.700&0.636&0.652~~~\\
~~~$\lambda_n$&$-$5.431&$-$5.435&$-$3.552&$-$3.543~~~\\
~~~$Q_{20}^p$ &410.08&411.31& 444.02 & 443.90~~~\\
~~~$Q_{20}^n$ &638.19&639.41& 788.32 & 786.63~~~\\
\end{tabular}
\end{ruledtabular}
\end{table}
Table~\ref{zrtab1} shows  the results of deformed calculations for
$^{102}$Zr and $^{110}$Zr with the same parameters as in spherical
calculations for  $^{120}$Sn displayed in Table~\ref{sntab2}. In
{\hfbtho}, deformed wave functions were expanded in the space of 20
stretched HO shells. The binding energies in {\hfbax} are greater by
about 110-140 keV than those of {\hfbtho}. This  is understandable
as {\hfbtho} with 20 shells also underestimates the binding energy
of $^{120}$Sn by about 150 keV~\cite{dobaczewski04}. In
Table~\ref{zrtab1} we only show neutron pairing (proton pairing
correlations in $^{110}$Zr vanish due to a deformed proton
subclosure at $Z$=40). It is gratifying to see that
 energies and
quadrupole moments of $^{102}$Zr and $^{110}$Zr  are very close in
{\hfbax} and {\hfbtho}, in spite of fairly  different
computational strategies  underlying these two codes.

\begin{table}[htb]
\caption{\label{zrtab2} Comparison between  {\hfbvan}
(second column)  and {\hfbax}  (third column) for $^{102}$Zr.
Calculation  parameters are the same as in Ref.~\cite{teran}, i.e.,
$V_0$=$-$170\,MeV\,fm$^3$, $\Omega_{\rm max}$=11/2, $R$=12\,fm, and
$N_{\rho}$=$N_{z}$=19. The results of {\hfbax} with the same density
functional but the standard box $R$=19.2\,fm and larger pairing cutoff
$\Omega_{\rm max}$=33/2 are displayed in the last column ({\hfbax}'). All energies
are in MeV. The mass rms radius $R_{rms}$ is in fm.}
\begin{ruledtabular}
\begin{tabular}{lccc}
& {\hfbvan} & {\hfbax}&{\hfbax}' ~~~~~ \\
 \hline
~~$E_{tot}$ &$-$859.61 &$-$859.19&$-$859.25~~~~~\\
~~$\lambda_n $&$-$5.46 &$-$5.47&$-5.45$~~~~~ \\
~~$\lambda_p$&$-$12.10 & $-$12.05&$-$12.06~~~~~\\
~~$\Delta_n$&  0.31& 0.28&0.43~~~~~\\
~~$\Delta_p$&  0.34& 0.37&0.40~~~~~\\
~~$R_{rms}$ &4.58& 4.58&4.58~~~~~\\
~~$\beta_2$ &0.431& 0.434&0.435~~~~~\\
\end{tabular}
\end{ruledtabular}
\end{table}
In order to compare {\hfbax} with the Vanderbilt lattice code
{\hfbvan}, we calculated the deformed nucleus $^{102}$Zr assuming
the same parameters as in Ref.~\cite{teran}, i.e., a fairly small
box radius $R$=12\,fm, coarse grid, and very low cutoff $\Omega_{\rm
max}$=11/2. The ground state of $^{102}$Zr is reflection symmetric;
thus, apart from the fact that the box of {\hfbvan} is twice as
large as that of {\hfbax}, the codes are supposed to produce the
same result. As seen in Table~\ref{zrtab2}, this is almost the case:
the difference for the binding energy is around 400\,keV. We believe
that this can be attributed to a slightly different structure of the
discretized positive-energy continuum in the two codes. Indeed the
average pairing gaps predicted in the two codes are  $\sim$30 keV
apart. To confirm this, we performed another calculation for
$^{102}$Zr in a larger space, i.e., $R$=19.2\,fm, $h$=0.6\,fm,  and
$\Omega_{\rm max}$=33/2. While the total energy is only weakly
affected, there is an appreciable increase in the pairing gaps. This
result, together with the discussion of $^{90}$Ni in
Sec.~\ref{nickl}, underlines the importance of using large boxes and
sizable pairing spaces for the description of neutron-rich systems.

%
% trim=l b r t
\begin{figure}[htb]
\centerline{\includegraphics[trim=0cm 0cm 0cm 0cm,width=0.4\textwidth,clip]{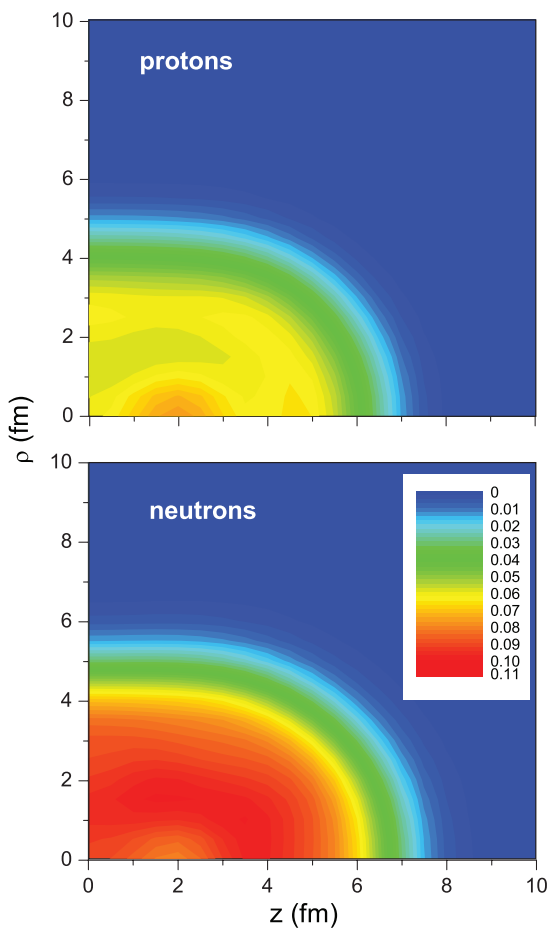}}
\caption{\label{zr110} (Color online) Contour plots of proton (top)
 and neutron (bottom) density distributions in the
 ($\rho,z$)-plane  for  the deformed
 ground state of $^{110}$Zr calculated in {\hfbax}.
The densities are    in nucleons/fm$^{-3}$.}
\end{figure}
\begin{figure}[htb]
\centerline{\includegraphics[trim=0cm 0cm 0cm 0cm,width=0.4\textwidth,clip]{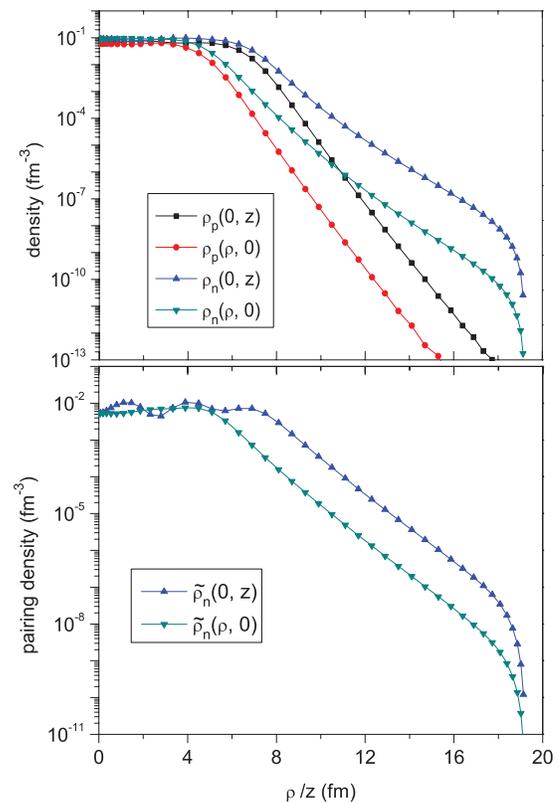}}
\caption{\label{zr110b}(Color online) Particle (top) and pairing
(bottom) ground-state densities in $^{110}$Zr along $\rho$=0 and
$z$=0.
 The size of the box is $R$=19.2\,fm.}
\end{figure}

Proton and neutron density distributions in  $^{110}$Zr are displayed in
Figs.~\ref{zr110}  (in two dimensions, to better show the deformed shape)
and \ref{zr110b} (in logarithmic scale, to better show the asymptotic behavior).
The appearance of the neutron skin beyond  the nuclear surface is clearly seen.
The density contours in Fig.~\ref{zr110} can be compared with the result
of  {\sc hfb-2d-lattice} shown in Fig.~6 of Ref.~\cite{blazkiewicz},
and there seems to appear a
good agreement between the two sets of calculations. In particular,
a small depression of density in the nuclear interior, due to shell effects,
is present in both cases.
Another interesting feature is the constancy of density diffuseness along the
nuclear surface.
%
% trim=l b r t
The asymptotic behavior of nuclear densities in Fig.~\ref{zr110b} is
consistent with general expectations \cite{dobaczewski96}, and the
ratio $\rho_n(0,z)/\rho_n(\rho=z,0)$  is roughly constant at large
distances. This indicates that densities are still  well deformed in the
region which is well beyond the nuclear surface.

\subsection{Large deformation limit: axial symmetric fission path of $^{240}$Pu}

The advantage of  coordinate-space calculations over HO expansion
methods  is apparent in the context of problems involving extreme
deformations, which require the use of huge oscillator spaces or
even a many-center HO basis. In this section, we study the axial,
reflection-symmetric fission path of $^{240}$Pu, which has been
investigated in many earlier works~\cite{DFTrev}. By carrying out
precise {\hfbax} calculations, one can assess the error on
 potential energy surfaces, energies of fission isomers, and fission
 barriers obtained in commonly used HFB codes employing an HO expansion technique.

The  HFB energy at a given mass quadrupole moment $Q_{20}=\langle \hat{Q}_{20}\rangle$
can be obtained by
minimizing the  routhian with a quadratic constraint~\cite{ring80}:
\begin{equation}
E'=E+C_q(\langle \hat{Q}_{20}\rangle -Q_{20})^2
\end{equation}
where
\begin{equation}
Q_{20}=2\pi\int\int\rho_{tot}(\rho , z)(2z^2-\rho^2)\rho {\rm d}\rho {\rm d}z,
\end{equation}
is the requested average  value of the mass quadrupole moment and
$C_q$ is the quadrupole stiffness constant.
%
% trim=l b r t
\begin{figure}[htb]
\centerline{\includegraphics[trim=0cm 0cm 0cm 0cm,width=0.4\textwidth,clip]{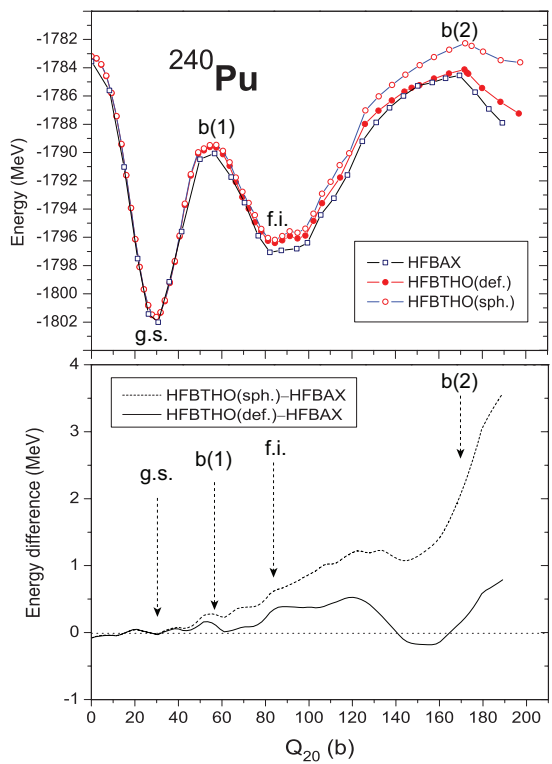}}
\caption{\label{pu240a}(Color online) Top: axial,
reflection-symmetric fission path of $^{240}$Pu (top) calculated
with {\hfbax} and {\hfbtho} (in a spherical and stretched HO basis).
Bottom: the difference between {\hfbtho} and {\hfbax} results
(normalized to zero at the ground-state configuration). The minima
and maxima of energy are marked: ground state, g.s.; first barrier,
b(1); fission isomer, f.i.; second barrier,  b(2).
 }
\end{figure}

The constrained {\hfbax}   calculations for $^{240}$Pu
 were performed in a box of
$R$=23.4\,fm and $h$=0.65\,fm, using a B-spline basis with $M$=11.
(For a similar mesh size, the  binding energy of $^{120}$Sn in
{\hfbax} agrees with {\hfbrad} within  50 keV.) The {\hfbtho}
calculations were carried out in a typical space of $N_{sh}$=20
spherical or stretched HO shells. The mixed pairing interaction is
used with the pairing strength adjusted as in Sec.~\ref{stablesph}
for $^{120}$Sn. The results are displayed in Fig.~\ref{pu240a}. The
spherical HO basis is unreliable for fission calculations, and the
quality of {\hfbtho} calculations  with a stretched basis
deteriorates gradually with deformation. That is, the energy error
on the first barrier and fission isomer is $\sim$100\,keV and
$\sim$400\,keV, respectively, and it reaches $\sim$500\,keV inside
the second barrier. These are significant corrections that can
impact calculated half-lives for spontaneous fission.

\section{Conclusions}\label{conclusions}

We developed a 2D coordinate space code  {\hfbax} using the
technique of basis splines. The high accuracy of the code has been
demonstrated by benchmarking it  against the multi-resolution
wavelet method, the HO basis expansion method, and the radial finite
difference method. The tests have been carried out for spherical and
very deformed configurations of stable and weakly bound nuclei.

A significant numerical speedup of the code makes it a useful tool
for nuclear structure calculations of exotic configurations, such as
halos and extremely elongated fissioning nuclei. Among the first
applications of {\hfbax}  planned are systematic studies of deformed
drip-line systems. The ability of the code to accommodate very large
angular momentum cutoffs is crucial in the context of nuclei with
large neutron skins and halos, in which the  high-$j$ continuum has
significant impact on pairing correlations \cite{Rotival}.

Other applications involve systematic studies of superdeformed
configurations and fission isomers. {\hfbax} will also provide
systematic energy corrections
at large deformations, essential for
HFB models of fission based on HO expansion. The
differences seen in Fig.~\ref{pu240a} are expected to appreciably
impact the predicted spontaneous fission half-lives.

In this work,  reflection-asymmetric  and triaxial deformations,
important in realistic fission calculations, have not been
investigated. We are currently developing a symmetry-free
coordinate-space 3D HFB solver based  on the multi-resolution
wavelet method. The  {\hfbax} code reported in this paper  will
provide crucial benchmark tests for the general-purpose wavelet HFB
framework.

\begin{acknowledgements}
This work was supported in part by the U.S.~Department of Energy
under Contract Nos.~DE-FG02-96ER40963 (University of Tennessee),
DE-AC05-00OR22725 with UT-Battelle, LLC (Oak Ridge National
Laboratory), DE-FG05-87ER40361 (Joint Institute for Heavy Ion
Research), and DE-FC02-07ER41457 with UNEDF SciDAC Collaboration.
Computational resources were provided by the National Center for
Computational Sciences at Oak Ridge and the National Energy Research
Scientific Computing Facility.
\end{acknowledgements}

\end{document}